\documentclass[aps,onecolumn,preprintnumbers,groupedaddress,nofootinbib]{revtex4}

\IfFileExists{srcltx.sty}{\usepackage[active]{srcltx}}

\begin{document}
\title{Cosmological and Astrophysical Implications of \\ Sterile Neutrinos}
\author{Kalliopi Petraki}
\affiliation{Department of Physics and Astronomy, University of California, Los Angeles, CA 90095-1547, USA}
\preprint{UCLA/09/TEP/58}

\begin{abstract}
\centering
\begin{minipage}{0.8\textwidth}
The discovery of neutrino masses implies the existence of new particles, the sterile neutrinos. These particles can have important implications for cosmology and astrophysics. A sterile neutrino with mass of a few keV can account for the dark matter of the universe. Its relic abundance can be produced via different mechanisms. A minimal extension of the Higgs sector of the Standard Model, with a gauge-singlet boson coupled to sterile neutrinos, can provide a consistent framework for the theory of neutrino masses, and can produce relic keV sterile neutrinos via decays of the singlet Higgs. This mechanism operates around the electroweak scale, and has interesting consequences for the electroweak phase transition. The resulting dark matter is ``colder" than the one produced via oscillations. This property changes the small-scale structure formation limits. Heavier sterile neutrinos can be produced in supernova cores and affect the thermal evolution of the star. Being short-lived, they decay inside the envelope and facilitate the energy transport from the core to the vicinity of the supernova shock. This enhances the probability for a successful explosion.
\end{minipage}
\end{abstract}

\maketitle

\section{Introduction}
Neutrino masses can be incorporated into the Standard Model (SM) by adding singlet fermions which play the role of right-handed neutrinos. The latter are allowed to have Yukawa couplings to the Higgs boson and the standard, left-handed neutrinos. The Yukawa couplings generate the Dirac masses for the neutrinos after the spontaneous symmetry breaking (SSB). In addition, the singlet fermions can have some Majorana masses. The SM Lagrangian extended to include the new states becomes:
\begin{equation}
\mathcal{L} = \mathcal{L_{_\mathrm{SM}}} + i \bar{N}_a {\partial}\hspace{-2mm}/ N_a
- \left( y_{\alpha a} \epsilon^{ij}  H^{\dag}_i \,   (\bar{L}_\alpha)_j  N_a
+ \frac{M_a}{2} \; \bar{N}_a^c N_a + h.c. \right)
\label{L}
\end{equation}
where $\alpha = {e, \, \mu, \, \tau}$ and $a = 1, ..., n$ runs over the gauge-singlet neutrino species.

The interplay between the Dirac and the Majorana mass terms, known as the seesaw mechanism, can accommodate the observed neutrino masses for a variety of choices. If the Majorana masses are large, the particles associated with the singlet fields are very heavy. However, if one or more Majorana masses are below the electroweak scale, the so-called sterile neutrinos appear among the low-energy degrees of freedom and can take part in a number of observable phenomena. Here we will discuss some of the implications of sterile neutrinos at two different mass scales for cosmology and astrophysics.

Light sterile neutrinos of mass of a few keV can be the cosmological dark matter~\citep{DW,Cool,sterile DM + inflation,sterile DM + Higgs,sterile DM + Higgs + EWPT}; the same particles can facilitate the formation of the first stars~\citep{reioniz}. Their production in a supernova can explain the pulsar kicks~\citep{pulsars}.

Heavier sterile neutrinos of mass $\sim$~200 MeV can be produced in the core of a hot proto-neutron star, and provide an efficient energy carrier, with important implications for the thermal evolution of the star~\citep{heavy_sterile_SN}.

\section{Light sterile neutrinos as dark matter}
Sterile neutrinos of mass of a few keV and small mixing are stable particles. If they are produced in the early universe, they will be part of the dark matter (DM) today. A relic abundance of keV sterile neutrinos can be produced in various ways. Oscillations of active into sterile neutrinos peak at a temperature around $T \sim 100$~MeV, and produce warm DM~\citep{DW}. However, in the presence of large lepton asymmetry, the low energy spectrum of active neutrinos in preferentially converted into sterile neutrinos, and the resulting DM spectrum is highly non-thermal with much lower average momentum~\citep{Cool}. Sterile neutrinos can also be produced via decays, if they couple to a scalar field. This field can play the role of inflaton~\citep{sterile DM + inflation}, or it can be part of an extended Higgs sector. In this case, sterile neutrinos are produced at the electroweak scale~\citep{sterile DM + Higgs} and they constitute colder DM than the relic sterile neutrinos produced via oscillations in the absence of lepton asymmetry~\citep{sterile DM + Higgs,sterile DM + Higgs + EWPT,SSS}. This scenario has important implications for the electroweak phase transition (EWPT)~\citep{sterile DM + Higgs + EWPT}.

\subsection{Production at the electroweak scale}
In the SM, fermions acquire their masses via the Higgs mechanism. The same mechanism can be responsible for the Majorana masses of sterile neutrinos. This only requires substituting the Majorana mass terms with a coupling of sterile neutrinos to a gauge-singlet scalar field $S$ which will also be subject to a scalar potential that allows for SSB
\begin{equation}
\mathcal{L} = \mathcal{L_{_\mathrm{SM}}} + i \bar{N}_a {\partial}\hspace{-2mm}/ N_a
- y_{\alpha a} \epsilon^{ij}  H^{\dag}_i \,   (\bar{L}_\alpha)_j  N_a - \frac{f_a}{2} \; S \; \bar{N}_a^c N_a -V(H,S) + \mathrm{h.c}.
\label{L w S}
\end{equation}

In this model, the Majorana masses are not a new arbitrary mass scale; they arise after SSB due to the coupling of sterile neutrinos to the singlet scalar:
$M = f \langle S \rangle$, where $\langle S \rangle$ is the singlet scalar vacuum expectation value (VEV). The new coupling opens a new channel of production of sterile neutrinos, namely via decays of the singlet scalar, $S \rightarrow N \, N$.

Requiring that the sterile neutrinos produced constitute all of the DM of the universe relates the mass of the DM particle with the scale of production. The relic abundance produced is~\citep{sterile DM + Higgs,sterile DM + Higgs + EWPT}
\begin{equation}
\Omega_{_N} \sim  0.2
\left( \frac{f}{10^{-8}}\right)^3
\left(\frac{\langle S \rangle}{m_S}\right)
\left(\frac{33}{\xi}\right)
\label{W}
\end{equation}
where $m_S$ is the singlet scalar mass, and the factor $\xi$ will be discussed below. Assuming that $m_S$ and $\langle S \rangle$ are of the same order, the Yukawa coupling of the DM sterile neutrino to the singlet scalar is fixed at $f \sim 10^{-8}$. This is low enough to keep sterile neutrinos always out of equilibrium. It also implies that a keV sterile neutrino produced via this mechanism can be the DM if the singlet scalar lives at the electroweak scale
$\langle S \rangle \sim m_S \sim 10^2 \ \mathrm{GeV}$~\citep{sterile DM + Higgs}. This has important consequences for the EWPT, and for the properties of DM.

In addition, the DM abundance produced via this mechanism does not depend on the mixing angle of sterile neutrinos with the active species. This is in sharp contrast with the production mechanisms that involve oscillations, and alleviates the tension induced by current X-ray bounds.

\subsubsection{The electroweak phase transition}

Since the singlet scalar lives at the electroweak scale, it will take part in the EWPT through its coupling to the SM Higgs $H$. The most general renormalizable scalar potential is
\begin{eqnarray}
V(H,S) = &-& \mu_H^2 |H|^2 - \frac{1}{2}\mu_S^2 S^2 + \frac{1}{6}\alpha S^3 + \omega |H|^2 S \nonumber \\
&+& \lambda_H |H|^4 + \frac{1}{4}\lambda_S S^4 + 2\lambda_{HS}|H|^2 S^2
\label{V}
\end{eqnarray}
The modified Higgs sector (\ref{V}) possesses several interesting features, and has been studied in detail by several authors (e.g.~\citet{EWPT}). It is a minimal extension of the SM, which, for some range of parameters, allows for a 1st order EWPT, provides a DM candidate, and has observable signatures for accelerator experiments. A strong 1st order EWPT opens the possibility for electroweak baryogenesis. The $S$ boson itself, if stable, can constitute the DM.

However, in the simplest version of the model where $S$ is a real field, these two features are mutually exclusive. A 1st order EWPT is made possible due to the presence of the odd power terms $\alpha S^3$ and $\omega |H|^2 S$ in (\ref{V}). These terms increase the asymmetry between the true and the false vacuum, but they also render the $S$ boson unstable and therefore not a viable DM candidate. On the other hand, if one imposes a discrete $Z_2$ symmetry on the singlet scalar, $S$ bosons become stable and their relic abundance can account for the DM, but this eliminates the prospects for electroweak baryogenesis in this model.

The above description is valid as long as the $S$ boson couples only to SM fields. However, if the singlet scalar couples to sterile neutrinos another possibility exists~\citep{sterile DM + Higgs + EWPT}. In the most general case, no $Z_2$ symmetry is present, and the EWPT can be 1st order. The  $S$ bosons decay into light sterile neutrinos. The latter can constitute the DM of the universe. The extended Higgs sector can be probed at the LHC~\citep{EWPT}, while the DM sterile neutrino can be discovered by X-ray observations of DM dominated galaxies (for recent limits see~\citet{Suzaku}).

\subsubsection{Dark-matter redshifting}
The sterile neutrinos produced at the electroweak scale constitute colder DM than the relic sterile neutrinos produced via oscillations, in the standard scenario~\citep{DW}. The thermal content of DM affects the clustering structure of the universe. The latter can be used to obtain limits on how warm dark matter can be. This property depends on the DM particle mass and on the production mechanism. The relevant limits for sterile neutrinos produced via decays of a singlet Higgs will thus be different than those quoted for sterile neutrinos produced via oscillations~\citep{sterile DM + Higgs}.

The production of sterile neutrinos via decays peaks at the latest time before the equilibrium density of $S$ bosons becomes thermally suppressed, i.e. at $T_\mathrm{prod} \sim m_S \sim 10^2 \ \mathrm{GeV}$. At this temperature, all of the SM degrees of freedom are thermally coupled. As the universe expands isentropically, the SM degrees of freedom decouple and release their entropy to the thermally coupled species. Due to their very weak coupling, the keV sterile neutrinos were never part of the thermal history, thus they do not receive any of the entropy released. At a later time they appear colder and more dilute in respect to the rest of the universe.

The dilution of the sterile-neutrino relic abundance since the time of production is characterized by the factor $\xi = \frac{g_*(T_\mathrm{prod})}{g_*(T_\mathrm{today})}$ (c.f. eq.~(\ref{W})), where $g_*(T)$ is the number of relativistic degrees of freedom when the universe is at temperature $T$. For production at the electroweak scale, $\xi = 110/3.3 \simeq 33$. Sterile neutrinos are produced at the electroweak scale with a non-thermal spectrum and average momentum lower than the thermal one at the same temperature,
$\left.\frac{\langle p_{_N} \rangle}{3.15 \, T}  \right|_\mathrm{prod} = 0.8$.
At later times their spectrum is redshifted by $\xi^{1/3}$ and their average momentum becomes
$\left.\frac{\langle p_{_N} \rangle}{3.15 \, T}  \right|_\mathrm{today} = 0.8 \, /\, \xi^{1/3} \simeq 0.2$.

\subsection{Small-scale structure formation}

The clustering structure of the universe at large scales can be reproduced equally well both in the cold dark matter (CDM) and in the warm dark matter (WDM) scenarios. However, at small scales there are several points of disagreement between CDM predictions and current observations of the subgalactic structure.
In particular, CDM simulations
produce a vastly larger number of DM subhalos, than the number of satellite galaxies observed~\citep{Via Lactea};
they overestimate the number of halos in low-density voids~\citep{voids};
they cannot account for the existence of pure-disk or disk-dominated galaxies~\citep{Pure-disk galaxies};
they lead to greater loss of angular momentum, due to gas condensation, than what observations suggest~\citep{Angular Momentum Problem};
they predict cuspy density profiles at the center of the galaxies~\citep{Via Lactea}, while observations favor cored profiles~\citep{Q_obs + cores v cusps}.

It is possible that these discrepancies will go away as simulations and observations improve. Indeed, some of the problems mentioned here can be remedied within the CDM scenario, by invoking complex astrophysical solutions which usually address only individual issues. However, it is also possible that the resolution lies outside the CDM paradigm. WDM intrinsically suppresses structure at small scales and is free altogether of the problems present in the CDM model.

The agreement of a particular WDM model with observations has to be tested with simulations. WDM simulations are intrinsically more complex than CDM, and given the variety of WDM candidates with different primordial power spectra, the task becomes formidable. It is still possible to gain some intuition about WDM candidates by employing quantities, such as the free-streaming length and the phase-space density, which capture some of the properties of WDM and can be compared with observations. This can also facilitate comparison among various dark-matter candidates, and yield important constraints.

\subsubsection{Free-streaming length}

Dark matter consists of collisionless particles which can stream out of overdense into underdense regions, and smooth the inhomogeneities. To the extent that this occurs, structures cannot grow. The scale below which density fluctuations are suppressed, formally the cutoff of the power spectrum of the density perturbations, is the free-streaming length, $\lambda_\mathrm{fs}$~\citep{lfs}
\begin{equation}
\lambda_\mathrm{fs} = \sqrt{\frac{\pi}{ G}} \ \frac{\left\langle \left(p^2/m^2\right)^{-1}\right\rangle^{-1/2}}{\rho^{1/2}}
\label{lfsdef}
\end{equation}

CDM has zero thermal velocities and thus zero free-streaming length. In this scenario there is no minimum scale at which clustering can occur. Non-zero free-streaming length results into suppression of structure at small scales $l<\lambda_\mathrm{fs}$.

Although comparing $\lambda_\mathrm{fs}$ with observations requires that the latter probe much smaller scales than currently possible, the free-streaming length can still serve as a useful parameter to compare various WDM models.

\subsubsection{Coarse-grained phase-space density}

The thermal content of DM today has two origins: the primordial thermal spectrum of the relic particles, and the heating that occurs during gravitational clustering. Our goal is to constrain the former.

The phase-space packing observed in the subgalactic structures sets an upper limit to the primordial thermal content of DM. It can be studied in terms of the coarse-grained phase-space density~\citep{Q defin}
\begin{equation}
Q \equiv \frac{\rho}{\left\langle p^2/m^2\right\rangle^{\frac{3}{2}}}
\label{Qdef}
\end{equation}

Q is an observable quantity today; in the non-relativistic limit $Q = \rho/ \langle v^2 \rangle^{3/2}$, and all of the information needed can be deduced from the galactic rotation curves, using observations of luminous matter. Recent observations of the Dwarf Spheroids yield $Q_\mathrm{obs} \sim  6 \cdot 10^{-6} - 6 \cdot 10^{-4} ~ \frac{M_\odot/\mathrm{pc}^3}{\mathrm{(km/s)^3}}$~\citep{Q_obs + cores v cusps}.

For collisionless particles and in the absence of self-gravity, $Q$ is also a Liouville invariant quantity~\citep{Q defin}. This allows us to compare the primordial with the observationally inferred value. Gravitational dynamics, however, can cause Q to decrease. The decrease in $Q$ reflects the entropy increase caused by dynamical heating during the non-linear gravitational clustering~\citep{Qdecrease-theory,Qdecrease-postulation}.

Since the phase-space density can only decrease, the primordial value $Q_p$ sets an upper limit to the dark-matter density $\rho$ today. In the CDM scenario, $Q_p\rightarrow \infty$, and there is no limit to the central density. For WDM, however, $Q_p$ is finite and halos form central cores rather than divergent cusps.

Phase-space considerations yield lower limits for the mass of DM candidates. Observations of luminous matter constrain how relativistic DM particles could have been at the of production, i.e. set an upper limit on the ratio $p/m$. The DM production mechanism informs us about how warm DM was at the time in respect to the rest of the universe, i.e. it provides the ratio $p/T$. Thus, for any production mechanism there is a minimum value for $m_{_\mathrm{DM}}$, below which DM would be too warm to be consistent with the phase-packing observed today.

\medskip

$\lambda_\mathrm{fs}$ and $Q_p$ for sterile neutrinos produced via oscillations and decays, are presented in table \ref{m limits}~\citep{lfs,Q defin,SSS}, together with a lower mass limit derived from phase-space density arguments.
\begin{table}[ht!]
  \centering
  \begin{tabular}{l|r|r|c}
    \hline
    \hline
    \parbox[c]{2.6cm}{sterile neutrino \\ production \\ mechanism}
    & \parbox[c]{1.7cm}{\centering \smallskip Free-streaming \\ length $\lambda_\mathrm{fs}$ \\ (kpc)}
    & \parbox[c]{2.6cm}{\centering Phase-space \\ density \\ $Q_p \ \left(\frac{M_\odot/\mathrm{pc}^3}{\rm{(km/s)^3}}\right)$}
    & \parbox[c]{2.8cm}{\centering sterile neutrino \\ lower mass limit \\ $m_{_N}^\mathrm{min}$} \\
    \hline
    \parbox[c]{2.5cm}{ non-resonant \\ oscillations }
    & $7   \,  \left(\frac{\mathrm{keV}}{m_{_N}}\right)$
    & $2.2   \cdot 10^{-5} \,  \left(\frac{m_{_N}}{\mathrm{keV}}\right)^3$
    & $0.76 \, \mathrm{keV} \left[\frac{Q_\mathrm{obs}}{10^{-5} \frac{M_\odot/\mathrm{pc}^3}{\mathrm{(km/s)^3}}}\right]^{\frac{1}{3}}$ \\
    \hline
    \parbox[c]{2.5cm}{resonant oscill. \\ (lepton asym.)}
    & $1.7  \,  \left(\frac{\mathrm{keV}}{m_{_N}}\right)$
    & $3.7 \cdot 10^{-3} \,  \left(\frac{m_{_N}}{\mathrm{keV}}\right)^4$
    & $0.23 \, \mathrm{keV} \left[\frac{Q_\mathrm{obs}}{10^{-5} \frac{M_\odot/\mathrm{pc}^3}{\mathrm{(km/s)^3}}}\right]^{\frac{1}{4}}$ \\
    \hline
    \parbox[c]{2.5cm}{ decays at the \\ EW scale }
    & $2    \,  \left(\frac{\mathrm{keV}}{m_{_N}}\right)$
    & $2.4 \cdot 10^{-4} \,  \left(\frac{m_{_N}}{\mathrm{keV}}\right)^3$
    & $0.35 \, \mathrm{keV} \left[\frac{Q_\mathrm{obs}}{10^{-5} \frac{M_\odot/\mathrm{pc}^3}{\mathrm{(km/s)^3}}}\right]^{\frac{1}{3}}$ \\
    \hline
    \hline
  \end{tabular}
  \caption{$\lambda_\mathrm{fs}$ and $Q_p$ for dark-matter sterile neutrinos produced via oscillations, in the absence and in the presence of lepton asymmetry, and via decays of a scalar at the electroweak scale. Phase-space density considerations yield a lower mass limit for each mechanism, given in the last column, for an average estimate of the observed phase-space density.}
  \label{m limits}
\end{table}

\section{Heavy sterile neutrinos in supernovae\label{SN}}

Sterile neutrinos with masses $\sim 0.2 \ \mathrm{GeV}$ and small mixing $\sin^2 \theta \sim 10^{-8} - 10^{-7}$, with either $\nu_\mu$ or $\nu_\tau$, could be produced in supernova (SN) cores and subsequently augment, via their decay, the energy transport from the core to the region around the stalled shock, thereby increasing the prospects for a core collapse supernova explosion. This scenario could be testable by observations of the neutrino signal from galactic supernovae~\citep*{heavy_sterile_SN}.

In a core-collapse SN, the gravitational energy of the progenitor star, $3 \times 10^{53}$~erg gets trapped in the core. At the bounce, this energy is released, producing a powerful shock. The shock ejects the stellar envelope and leaves behind a compact remnant, a neutron star or a black hole. The optical output of the explosion accounts for as much as $10^{51} - 10^{52}$~erg. As confirmed by observations of SN1987A almost all of the gravitational energy is released in active neutrinos.

However, current simulations of the core collapse have not yet robustly established the physics that governs the propagation of the shock. 1d and 2d simulations fail to reproduce a successful explosion, the reason being that dissociation of iron nuclei absorbs the energy of the shock as it propagates. Multidimensional hydrodynamic simulations with full implementation of transport equations are still in the early stages of development. While it is possible that they will eventually succeed in reproducing successful explosions~\citep{SN simulations}, it is also possible that the energy transport in a SN involves new physics.

In particular, heavy sterile neutrinos can provide an excellent energy carrier. They can be produced in the core of a hot proto-neutron star by the standard electroweak interactions. If they mix weakly with the active species, they are not trapped and escape freely from the SN core. Being heavy, they drain only a small portion of the total energy of the core, while most of the energy is still carried away by the active neutrinos at the bounce, in accordance to the standard scenario. Nevertheless, the small amount of energy carried by the sterile neutrinos can be significant for the propagation of the shock, if deposited in its vicinity. Heavy sterile neutrinos are short-lived, thus they decay soon after their production, depositing their energy within the envelope of the star.

The feasibility of the above scenario depends on the mass and the mixing angle of sterile neutrinos. We focus on sterile neutrinos with masses $145-250$~MeV and vacuum mixing $\theta^2 \sim 10^{-8} - 10^{-7}$. The current constraints~\citep{limits:nu_mu mixing} allow this parameter range for mixing with both $\nu_\mu$ and $\nu_\tau$.  For mixing only with $\nu_\tau$ an even broader range is allowed~\citep{limits:nu_tau mixing}.

The flux of heavy sterile neutrinos from a SN depends strongly on the temperature in the core. SN cores typically reach peak temperatures $T \sim 35$~MeV, at $R \sim 10-20$~km, for a time interval of a few seconds. Numerical calculations show that the sterile neutrino burst will last 1-5~s and drain total energy from a supernova core $E_s \sim 10^{51} - 10^{52} \mathrm{erg}$~\citep{heavy_sterile_SN}.
This result is reproduced for specific models of SN cores from 1d simulations~\citep{SN simul 1d -Pons,SN simul 1d -Buras}. This amount of energy is of the order of the optical output of a SN.

The dominant decay mode of $\sim$~200~MeV sterile neutrinos is
\begin{equation}
N_s \rightarrow \nu_{\mu,\tau} + \pi^0 \rightarrow \nu_{\mu,\tau} + 2\gamma
\end{equation}
Their decay time is $\tau_{_d} \sim 0.1 \ s$~\citep{sterile decay}. Emitted from the core with $\gamma_s \sim 1.5$, sterile neutrinos deposit their energy at distances $R_s \sim 10^9$~cm, i.e. outside the neutron star but inside the envelope.

The $\gamma$-ray photons produced in the decays quickly thermalize with matter, and generate a pressure gradient that can affect the standard picture of the gravitational collapse. The amount of energy they carry can increase the entropy by a few units of Boltzmann's constant per baryon. In this event, nuclei in nuclear statistical equilibrium would be melted \citep{FMW} and at least some of the photo-dissociation burden on the shock could be alleviated. The prospects for an explosion likely would be enhanced in this scenario, even in a simplistic 1d supernova model~\citep{heavy_sterile_SN}.

The sterile neutrino decay will also produce $\nu_\mu$s or $\nu_\tau$s with energies around 80~MeV. Although very energetic, these neutrinos likely will escape from the SN, and provide a signature for the scenario presented here. In the event of a galactic SN, experiments could detect a short (1--5 s) burst of energetic $\nu_{\mu,\tau}$ from heavy sterile neutrino decays, followed by a longer (10--15~s) burst of $\sim 15$~MeV neutrinos. The detection probability of the energetic neutrinos is larger than that of ordinary neutrinos. However, they constitute only a small portion of the total number of neutrinos emitted. Their time-integrated flux will be a small fraction of the total neutrino flux $F_{80}/F_{\mathrm{total}} \sim 2\%$~\citep{heavy_sterile_SN}.

\section{Conclusions}
Sterile neutrinos arise as a compelling explanation of the neutrino masses. But they also have important implications for cosmology and astrophysics which can be employed to inform the direction of current and future experimental searches.

Sterile neutrinos of a few keV can be the dark matter. Their non-negligible primordial thermal content can alleviate the discrepancies between the CDM scenario and current observations of the subgalactic structure. Their relic abundance can be produced in the early universe by various mechanisms, each of which possesses different clustering properties and is thus subject to different limits from small-scale structure considerations. Dark-matter sterile neutrinos can facilitate the star formation and the reionization of the universe. The same particles can be produced in supernovae and explain the pulsar velocities.

Heavy and unstable sterile neutrinos can play an important role in various astrophysical phenomena. Sterile neutrinos with mass $\sim$~200~MeV, and mixing $\theta^2 \sim 3 \cdot 10^{-8}$ can be produced in the core of a hot proto-neutron star and enhance the energy transport within the star. This is very important for the propagation of the shock after the gravitational collapse, and ultimately the fate of the star.

\acknowledgements
This work was supported in part by the DOE grant DE-FG03-91ER40662 and by the NASA ATP grant NNX08AL48G.

\end{document}